\documentclass[preprint,prl,aps]{revtex4-2}
\usepackage{amsmath, amssymb, physics}
\usepackage{geometry}
\usepackage{graphicx}
\usepackage{hyperref}

\def\mnras{Mon. Not. Roy. Astron. Soc.}
\def\apjl{Astrophys. J. Lett.}
\def\apj{Astrophys. J.}
\def\aapr{Astron. Astrophys. Rev.}
\def\araa{Annu. Rev. Astron. Astrophys.}
\def\nature{Nature}
\def\pre{Phys. Rev. E}

\newcommand{\be}{\begin{equation}}
\newcommand{\ee}{\end{equation}}
\newcommand{\nn}{\mbox{} \nonumber \\ \mbox{} }
\newcommand{\ba}{\begin{eqnarray}}
\newcommand{\ea}{\end{eqnarray}}
\newcommand{\om}{\omega}
\newcommand{\Alfven}{ Alfv\'{e}n }

\newcommand\eg{{\it{{e.g., }}}}

\newcommand{\Lf}{{Lorentz factor}}

\newcommand{\Bf}{{magnetic field}}

\newcommand{\Ef}{{electric  field}}

\newcommand{\NSs}{{neutron stars}}
\newcommand{\EM}{electromagnetic}

\newcommand{\mss}{magnetospheres}

\begin{document}

\title{Particle dynamics in nonlinear electromagnetic  waves: chaos onset, diffusive heating,  and wave surfing}
\author{Maxim Lyutikov \\
Department of Physics and Astronomy, Purdue University, \\ 525 Northwestern Avenue, West Lafayette, IN 47907-2036 }

\begin{abstract}
 We investigate the dynamics of charged particles interacting with ultra-intense electromagnetic X-modes in strongly magnetized plasmas. We demonstrate that particle motion becomes chaotic for relative wave intensities $\delta = B_w/B_0 \gtrsim 0.25$  ({\it not}  above the field reversal threshold $\delta \geq 1$). The transition to chaos occurs via the Chirikov resonance overlap mechanism and the related destruction of Kolmogorov-Arnold-Moser (KAM) tori.  The maximum Lyapunov exponent increases logarithmically with $\delta$, even though the unmagnetized $\delta \to \infty$ limit is strictly integrable.  In the $\delta \gg 1$ regime, incomplete re-laminarization of the phase space flow leads to two distinct populations: (i) the majority of particles undergoing stochastic diffusion, and (ii) a fraction of particles that become phase-locked with the wave, experiencing macroscopic intermittent surfing (L\'evy flights).  The 1D Particle-In-Cell simulations using the EPOCH code in the highly magnetized ($\sigma \gg 1$) and under-dense  regime are generally consistent with the Hamiltonian  single-particle theory. The dissipation fraction  of the initial \EM\ energy  remains   mild. 
\end{abstract}

\maketitle

\section{Introduction}
The physics of ultra-strong laser-matter interaction became a forefront research topic in relativistic plasma astrophysics, initiated by the meteoritic developments over the last years in the field of mysterious Fast Radio Bursts (FRBs). {The plasma-physics challengers are enormous: how to produce and propagate ultra-intense millisecond-long radio bursts from $\sim $ halfway across the visible Universe} \citep{2007Sci...318..777L,2016MNRAS.462..941L,2022A&ARv..30....2P,2019ARA&A..57..417C,2016Natur.531..202S}.

Observations of correlated radio and X-ray bursts from ultra-magnetized \NSs\ \citep{2020Natur.587...54C,2021NatAs...5..372R,2020Natur.587...59B,2020ApJ...898L..29M,2021NatAs.tmp...54L} 
establishes the FRB-magnetar connection. If emission originates in the \mss, the laser non-linearity parameter in these settings can be as high as staggering $a_0 \sim 10^9$.


An important difference from the laboratory/laser plasma interaction is that in astrophysical setting the  external magnetic guide can be dominant, so that 
 the cyclotron frequency associated with the guide field is  comparable or larger   than the plasma frequency $\om_p$ and the radiation frequency $\om$\citep{1979PhFl...22.1089K,2013PhRvS..16i0701G}. Thus, the  cyclotron frequencies associated with the guide  \Bf, the fluctuating \Bf,  the plasma frequency,  and of the  wave frequency  itself can be  all comparable.

Two somewhat different plasma physics issues are (i) generation of FRBs; (ii) escape of nonlinear waves propagating through  the decreasing \Bf. As for the origin,
the most compelling model, in our view, is the "Solar flare paradigm": generation of coherent radio emission during magnetospheric reconnection events \citep{2002ApJ...580L..65L,2003MNRAS.346..540L,2006APS..APR.X3003L,2020arXiv200505093L}.

 As for the escape, in a series of works \cite{2023ApJ...959...34B,2026arXiv260610189B} it was argued that for the X-mode  with the amplitude larger than that of the background \Bf,  $ \delta \equiv B_w /B_0 \geq 1$,
at phases when  the  instantaneous  \Ef\ exceeds the total \Bf\ (guide minus fluctuating), the plasma drift speed  approaches c, so it experiences ultra-relativistic acceleration.

 Besides a number of macroscopic issues that may  allow waves to escape (most important are ponderomotive effects - ponderomotive broom - and bulk motion effects, 
\cite{2024MNRAS.529.2180L,2026arXiv260609417L}), the statement that instantaneous $|E| > |B|$ leads to fast acceleration   is incorrect on several  grounds.  The counter-arguments are many.
 Any superluminal wave, e.g. the basic EM wave in plasma  with $\om^2 = \om_p^2 + (kc )^2$,  has $|E|  > |B|$ - particles are not accelerated. 
Even in vacuum, only in {\it constant} crossed  electric and magnetic fields  the drift velocity becomes superluminal for $|E| >|B|$. Generally, as the name signifies, drifts are averaging over some fast periodic motion; instantaneous $|E| > |B|$ does not mean much.
 
 Second,  the MHD approximation is inappropriate - as particle dynamics is sensitive  to kinetic/phase-depended effects  (this is the main point  of the present paper). 

Finally,  in plasma, where  the current reacts both to electric and magnetic fields  in a collective manner, the vacuum logic  should be used with caution.

On the other hand, intense   plasma-EM wave energy exchanges were indeed observed in PIC simulations \citep[\eg][]{2025PhRvL.134c5201V}. In this work we look at the details of the nonlinear \EM\  wave plasma energy exchange.   It is not   E-cross-B drift:  there are intricate phase trajectories, resulting in  kinetic/phase dependent, chaotic behavior.

Historically,  the transition to global stochasticity and chaotic particle motion under the influence of coherent electromagnetic waves and background magnetic fields,  the  underlying theory of resonance overlap, and transition to  deterministic chaos were addressed  in Ref. \cite{Chirikov1979}, and extensively applied to the stochastic heating of magnetized plasma particles in Ref. \cite{Karney1978}.

\section{Field Configuration:  nonlinear X-mode propagating perpendicular to the \Bf}
Consider a relativistic particle  interacting with a uniform background magnetic field and   linearly polarized electromagnetic X-mode propagating perpendicular to the \Bf.
(The oblique guide field configuration, with non-zero $B_x$,  can be  transformed into a pure transverse  one  via a  Lorentz transformation.)
The wave propagates along the $x$-axis with the electric field polarized along the $z$-axis. The background magnetic field is oriented along the $y$-axis.


The wave has an arbitrary phase velocity characterized by $\beta = {\omega}/{  k}$.
The vector potential $\mathbf{A}$, chosen in the Coulomb gauge ($\Phi = 0$) for the wave and Landau  gauge for the guide field, is aligned along the $z$-axis:
\begin{equation}
\mathbf{A} = \mathbf{A}_{ext} + \mathbf{A}_w = -  \left( B_0 x + a_0 \sin(\omega t - kx) \right) \hat{\mathbf{z}}.
\label{AA} 
\end{equation}
where $a_0 = E_w/\omega$ is the laser intensity parameter and $B_0$ is the guide field along  $\hat{\mathbf{y}} $ direction.
(We set the speed of light, charge, and mass to unity, $c=1, e=1, m=1$.)

\section{The Hamilton-Jacobi Equation for particle motion in nonlinear X-mode}
The relativistic Hamiltonian for the particle is:
\begin{equation}
H = \sqrt{1 + P_x^2 + P_y^2 + \left( P_z - A_z \right)^2},
\end{equation}
where $\mathbf{P}$ is the canonical momentum. Since the Hamiltonian does not depend  on $y$ and $z$, the canonical momenta $P_y$ and $P_z$ are conserved quantities. For simplicity, we can assume $P_y = P_z = 0$ without loss of generality for motion strictly in the $x$-$z$ plane.

The Hamilton-Jacobi (HJ) equation is obtained by substituting $H = -{\partial_t S}$ and $\mathbf{P} = \nabla S$
 \footnote{Alternatively,  one  could start with the covariant Hamilton-Jacobi equation for a charged particle in an electromagnetic field \citep{LLII}
\begin{equation}
g^{\mu \nu} \left( \partial _\mu S + A_\mu \right) \left(  \partial _\nu S + A_\nu \right) + 1 =0
\end{equation}
}:
\begin{equation}
\left( \frac{\partial S}{\partial t} \right)^2 = 1 + \left( \frac{\partial S}{\partial x} \right)^2 + \left( B_0 x +   a_0\sin(\omega t - kx) \right)^2.
\end{equation}

We introduce the traveling wave phase variable $\xi$ and an auxiliary time variable $\tau$:
\begin{equation}
\xi = \omega t - kx, \quad \tau = t.
\end{equation}
The partial derivatives transform as:
\begin{align}
\frac{\partial}{\partial t} &= \omega \frac{\partial}{\partial \xi} + \frac{\partial}{\partial \tau}, \\
\frac{\partial}{\partial x} &= -k \frac{\partial}{\partial \xi}.
\end{align}
For the transformed action $\bar{S}(\xi, \tau)$, the HJ equation becomes:
\ba &&
\omega^2(1 - n^2) \left( \frac{\partial \bar{S}}{\partial \xi} \right)^2 + 2\omega \frac{\partial \bar{S}}{\partial \xi} \frac{\partial \bar{S}}{\partial \tau} + \left( \frac{\partial \bar{S}}{\partial \tau} \right)^2 = 1 + \left( B_0 \frac{\omega \tau - \xi}{k} +   a_0\sin \xi \right)^2.
\nn &&
n= \frac{k}{\om}= \frac{1}{\beta}
\label{HJ}
\ea
This is the  Hamilton-Jacobi equation for the action $\bar{S}$, governing the particle dynamics in nonlinear X-mode propagating across \Bf. It depends on  the phase speed of the wave $n$, guide \Bf\ $ B_0 \equiv \om_c$, and wave intensity $a_0$.

Eq. (\ref{HJ}) demonstrates, that the luminal,  sub- and superluminal cases are qualitatively different, changing from parabolic to either
elliptical or hyperbolic regimes depending on the magnitude of $n$. In what follows, we treat these cases separately, see Supplemental Material for the non-luminal case.  

\section{Vacuum Regime ($n = 1$)}
In a vacuum, the  HJ equation  (\ref{HJ}) simplifies 
\begin{equation}
2 \omega \frac{\partial \bar{S}}{\partial \xi} \frac{\partial \bar{S}}{\partial \tau} + \left( \frac{\partial \bar{S}}{\partial \tau} \right)^2 = 1 + \left( B_0 \frac{\omega \tau - \xi}{k} +   a_0\sin \xi \right)^2.
\end{equation}

In the Supplemental Material, for consistency, we reproduce two integrable limits, zero guide field (figure-8) and zero wave intensity (cyclotron motion) using HJ approach.

\subsection{Vacuum X-mode with external magnetic field:  chaos onset}

In what follows, we  characterize the relative strength of the fields by the parameter 
\ba &&
 \delta = \frac{B_w}{B_0} = a_0  n \frac{\om}{\om_c} =  a_0  n  \tilde{\omega}  \to  a_0    \tilde{\omega} , \, \mbox{for $n=1$}
 \nn &&
 \om_c = \frac{e B_0}{m c}
 \nn && 
 \Omega_c = \frac{ \om_c}{\gamma}
 \nn &&
 \tilde{\omega} = \frac{\omega}{\omega_c}
 \ea
   ($n$ above is the refractive index; we also note a relation $a_0 = \delta   \sqrt{\sigma} (\om_p /\om)$).

    The system is integrable in the limit $\delta =0$ (cyclotron motion in constant \Bf), and  $\delta = \infty$ (\EM\ wave with no guide field). We are interested in particle dynamics in the intermediate regime of finite $\delta$.  

We chose to start from dominant guide field; see Supplemental Material for expansion near the limit of pure \EM\ wave. The introduction of the background magnetic field $B_0$  introduces $\tau$ into the Hamiltonian: this  breaks the symmetry that conserved $\mathcal{E}$, destroying the fundamental light-cone invariant $\Lambda$ (defined in the Supplemental Material).

Let us normalize the proper time of the particle using the rest cyclotron frequency $\omega_c $, defining the dimensionless proper time $s$ such that $dt = ({\gamma}/ {\omega_c} ) ds$. 
The Lorentz force gives the equations of motion for the momentum components:
\begin{align}
\frac{dp_x}{ds} &= -p_z (1 -  \delta \cos\xi), \\
\frac{dp_z}{ds} &= \gamma  \delta \cos\xi + p_x (1 -  \delta \cos\xi) = p_x + (\gamma - p_x)  \delta \cos\xi,
\end{align}

The energy evolution follows from
\be
\frac{d\gamma}{ds} = p_z  \delta \cos\xi.
\ee

The light-cone variable $\Lambda $ now evolves according to 
\begin{equation}
\frac{d\Lambda}{ds} = \frac{d\gamma}{ds} - \frac{dp_x}{ds} = p_z  \delta \cos\xi + p_z (1 -  \delta \cos\xi) = p_z.
\end{equation}
The evolution of the wave phase $\xi $ with respect to the proper time is:
\begin{equation}
\frac{d\xi}{ds} = \frac{\omega}{\omega_c} \gamma - \frac{k}{\omega_c} p_x = \tilde{\omega} (\gamma - p_x) = \tilde{\omega} \Lambda,
\end{equation}

Using the mass-shell condition $\gamma^2 - p_x^2 = 1 + p_z^2$, we can express $p_x$  in terms of $\Lambda$ and $p_z$: 
\be
p_x = \frac{1 - \Lambda^2 + p_z^2}{2\Lambda}.
\ee

 Substituting this into the $p_z$ equation yields a  closed  system that describes  the nonlinear dynamics:
\begin{align}
\frac{d\xi}{ds} &= \tilde{\omega} \Lambda, \\
\frac{d\Lambda}{ds} &= p_z, \\
\frac{dp_z}{ds} &= \frac{1 - \Lambda^2 + p_z^2}{2\Lambda} +  \boxed{ \delta \Lambda \cos\xi}.
\label{Main1}
\end{align}
This   demonstrates how the external magnetic field breaks the integrability. When $ \delta = 0$, the system is perfectly integrable. For finite $ \delta$, the boxed   term  leads to chaotic phase space trajectories.

\subsection{Chirikov Threshold and Lyapunov Exponents}
According to the Kolmogorov-Arnold-Moser (KAM) theorem \cite{Kolmogorov1954,1992AnRFM..24..145A}, for sufficiently weak wave perturbations, unbroken invariant KAM tori persist in the phase space. In a 1.5-degree-of-freedom non-autonomous system, these invariant curves act as absolute topological barriers that  compartmentalize the phase space, isolating adjacent cyclotron resonances and strictly forbidding unbounded macroscopic energy diffusion. 

The Chirikov resonance overlap criterion \cite{Chirikov1979} provides a way  to estimate the perturbation amplitude required to destroy the last isolating KAM torus. It dictates that global stochasticity emerges when the full width of a resonance island exceeds the energy separation between adjacent resonances. Defining the overlap parameter as 
\be
K = \frac{\Delta \gamma_{\text{island}}}{(\delta \gamma)},
\label{K}
\ee
the onset of global chaos --- coinciding with the topological breakdown of the isolating KAM barriers into percolating fractal cantori --- occurs at $K \ge 1$. Substituting the expressions for the island width and resonance separation, we obtain:
\begin{equation}
K =
4 \sqrt{a_0 \gamma_n} \tilde{\omega}^{3/2} \ge 1
\end{equation}
Squaring this condition to solve for the critical normalized wave amplitude $a_0^{\text{crit}}$ yields:
\begin{equation}
16 a_0^{\text{crit}} \gamma_n \tilde{\omega}^3 \approx 1 \implies a_0^{\text{crit}} \approx \frac{1}{16 \gamma_n \tilde{\omega}^3}
\end{equation}
We arrive at the  analytical threshold for the laser amplitude required to globally stochastize the phase space:
\begin{equation}
  a_0^{\text{crit}} \approx \frac{1}{16 \gamma_n \tilde{\omega}^3}
\label{A0crit}
\end{equation}

Deep inside the highly stochastic domain ($K \gg 1$), the dynamics can be locally modeled using the Chirikov standard map. For the standard map with stochastic parameter $K_s \propto a_0 \tilde{\omega}^3$, the trajectory divergence (Lyapunov exponent) per discrete mapping step is $\lambda_{\text{step}} \approx \ln(K_s/2)$. The discrete step maps to one unperturbed cyclotron period $T_c = 2\pi/\Omega_c$. The continuous wave phase $\xi$ advanced during one such cyclotron period is $\Delta \xi \approx \omega T_c = 2\pi \omega / \Omega_c$. 

To compute the  Lyapunov exponent $\lambda_{\max}$ with respect to the  wave phase $\xi$, we re-scale the discrete step divergence:
\begin{equation}
\lambda_{\max} \approx \frac{\lambda_{\text{step}}}{\Delta \xi} \approx \frac{\Omega_c}{2\pi \omega} \ln \left( \frac{K_s}{2} \right)
\end{equation}
Substituting the  scaling for the standard map parameter $K_s = 2 c_1  a_0 \tilde{\omega}^3$ (where $c_1$ is a proportionality constant of order unity), we obtain the final analytical expression for the maximal Lyapunov exponent:
\begin{equation}
\lambda_{\max} \approx \frac{\Omega_c}{2\pi \omega} \ln \left[ c_1   a_0 \tilde{\omega}^3 \right]
\label{lambdamax} 
\end{equation}

\subsection{Re-laminarization and macroscopic intermittent surfing ($ \delta \gg 1$)}
For extremely intense wave fields ($ \delta \gg 1$), the chaotic phase space transitions into a temporary re-laminarized state, 
see also Supplemental Material.
In this limit, the intense wave field {\it occasionally} surf-accelerates the particle to ultra-relativistic energies. 


To determine the threshold for this surfing phase, we examine the evolution of the light-cone invariant $\Lambda = \gamma - p_x$. In the highly nonlinear regime ($ \delta \gg 1$), the transverse momentum is dominated by the wave, yielding $dp_z/d\xi \approx ( \delta/\tilde{\omega}) \cos\xi$, which integrates to $p_z \approx ( \delta/\tilde{\omega}) \sin\xi$. The variation of the invariant with respect to the wave phase is then given by:
\begin{equation}
    \frac{d\Lambda}{d\xi} = \frac{p_z}{\tilde{\omega} \Lambda} \approx \frac{ \delta}{\tilde{\omega}^2 \Lambda} \sin\xi.
\end{equation}
Integrating this equation from an initial phase $\xi_0$ gives:
\begin{equation}
    \Lambda^2(\xi) \approx \Lambda_0^2 - \frac{2  \delta}{\tilde{\omega}^2} (\cos\xi - \cos\xi_0).
\end{equation}

The maximum deviation of $\Lambda^2$ occurs for particles injected at $\xi_0 = \pi$ and evaluated at $\xi = 0$, giving a maximal reduction of $\Delta(\Lambda^2) = 4 \delta / \tilde{\omega}^2$. 
For the particle to enter this macroscopic surfing phase, the wave must be strong enough to overcome the magnetic field's tendency to phase-wrap the trajectory, driving $\Lambda$  close to zero. As $\Lambda \to 0$, the energy $\gamma \approx \frac{1+p_z^2}{2\Lambda} \to \infty$ and the effective cyclotron frequency $\Omega_c \to 0$, breaking the resonance condition. Setting $\Lambda_{min}^2 = \Lambda_0^2 - \frac{4 \delta}{\tilde{\omega}^2} = 0$ yields the  threshold parameter $ \delta_{th}$ for the onset of surfing:
\begin{equation}
 \delta_{th} = \frac{\tilde{\omega}^2 \Lambda_0^2}{4}.
 \label{deltath}
\end{equation}

For a given initial injection energy $\gamma_0$ (assuming purely transverse initial momentum, $p_{x,0} = 0$, such that $\Lambda_0 \approx \gamma_0$), we can similarly express this as a threshold wave amplitude $a_0^{\text{surf}}$:
\begin{equation}
  a_0^{\text{surf}} \approx \frac{1}{4} \tilde{\omega}^2 \gamma_0^2
\label{a0surf}
\end{equation}

It is instructive to examine the relation between the parameter $\delta_{th}$ (Eq. \ref{deltath}) and the amplitude $a_0^{\text{surf}}$ (Eq. \ref{a0surf}). By definition, the relative intensity is $\delta = a_0 \tilde{\omega}$ (assuming a refractive index $n=1$), meaning the onset of surfing physically requires $\delta_{th} = a_0^{\text{surf}} \tilde{\omega}$. The two threshold expressions are closely related and characterize the identical physical boundary, with Eq. (\ref{a0surf}) often functioning as a practical measure in regimes where $\tilde{\omega} \sim 1$.

The relation between the onset of global stochasticity ($  a_0^{\text{crit}}$) and the onset of macroscopic surfing ($  a_0^{\text{surf}}$) is revealing. Combining Eq. (\ref{A0crit}) and (\ref{a0surf}), the ratio of the surfing threshold to the global chaos threshold is:
\begin{equation}
\frac{  a_0^{\text{surf}}}{  a_0^{\text{crit}}} \approx 4 \gamma_0^2 \gamma_n \tilde{\omega}^5
\label{A0ratio}
\end{equation}
(Here, it is important to distinguish between $\gamma_0$ and $\gamma_n$: the parameter $\gamma_0$ represents the actual, physical initial injection energy of the particle entering the system, whereas $\gamma_n$ is the mathematically defined resonant energy corresponding to the $n$-th cyclotron harmonic (i.e., the energy where $\omega - k v_x = n \Omega_c$).)

Eq. (\ref{A0ratio}) demonstrates that for particles injected near the fundamental resonance limit ($\omega \approx \omega_c$) with moderately low non-relativistic energies ($\gamma_0 \approx \gamma_n \approx 1$), this ratio simplifies to  $4$. This theoretically proves that the driving laser amplitude required to punch through the chaotic sea and reach the macroscopically re-laminarized surfing state is precisely four times larger than the threshold amplitude required to induce global chaos in the first place ($  a_0^{\text{surf}} \approx 4   a_0^{\text{crit}}$). This ratio perfectly aligns with the numerically observed regimes in Figures \ref{fig:poincare_deltas} and \ref{fig:lyapunov}, where chaotic stochasticity begins at $\delta \approx 0.0625$ and surfing strictly requires $\delta \ge 0.25$.

When $ \delta >  \delta_{th}$ (or correspondingly $a_0 > a_0^{\text{surf}}$), the external magnetic field acts merely as a negligible adiabatic perturbation, and the particle is temporarily decoupled from cyclotron resonances, bypassing discrete resonance structures and traversing an enormous macroscopic trajectory. As the particle becomes phase-locked in this surfing regime, the macroscopic motion forms a smooth, parabolic pathway in the laboratory frame, driven by the persistent ponderomotive surfing acceleration (see Supplemental Material). However, this {\it re-laminarization is never complete/permanent}. The residual magnetic field eventually bends the trajectory over exceptionally long times, causing the particle to dephase.
\subsection{Numerical Investigation of Phase Space Topology}

\begin{figure}[ht!]
    \centering
    \includegraphics[width=0.9\textwidth]{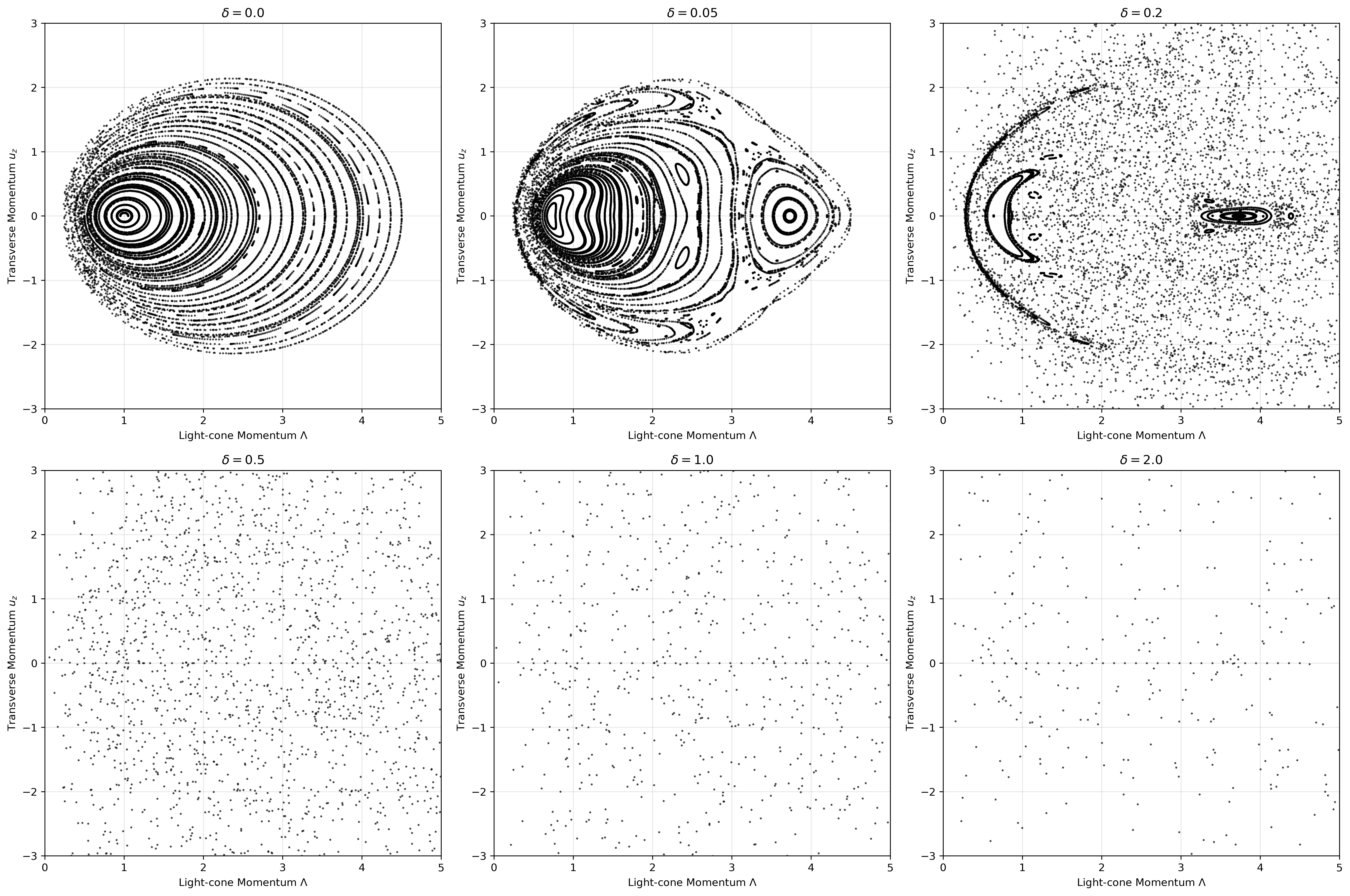}
    \caption{Poincar\'e maps illustrating the phase space topology for varying relative wave intensities $ \delta$. The transition flows from  integrable orbits ($\delta=0.0$) and near-integrable KAM tori ($\delta=0.05$) to overlapping islands ($\delta=0.2$), into global chaos ($\delta=0.5$ and $\delta=1.0$), and ultimately to intermittent macroscopic surfing ($\delta=2.0$).}
    \label{fig:poincare_deltas}
\end{figure}

To  visualize how the dynamical phase space evolves with the relative wave intensity, we numerically integrate the system  (\ref{Main1}) to generate Poincar\'e maps across four distinct regimes of $ \delta$, taking stroboscopic cross-sections at $\xi = 2\pi m$, Fig. \ref{fig:poincare_deltas} 

The progression captured in Figure \ref{fig:poincare_deltas} is characterized by the following regimes:
\begin{enumerate}
    \item   {Unperturbed Integrable System ($\delta = 0.0$):} The phase space is perfectly foliated by smooth, continuous invariant curves. No chaotic dynamics are present, and the particle exhibits strictly periodic, regular cyclotron gyration.
    \item   {Weak Perturbation ($ \delta = 0.05$):} The Kolmogorov-Arnold-Moser (KAM) tori strictly dominate the phase space. The wave creates very thin resonance islands at integer values of $\Lambda$ (for $\tilde{\omega}=1$). The Chirikov resonance overlap parameter is $K \ll 1$, meaning particles cannot wander between resonances. The phase space is highly structured and regular.
    \item   {Transition to Chaos ($ \delta = 0.2$):} The resonance islands widen significantly (island half-width $\Delta I \propto \sqrt{ \delta}$). The separatrix layers begin to overlap ($K \sim 1$). A "chaotic sea" forms between the primary islands, constituting structural phase noise.
    \item   {Strong Chaos ($ \delta = 0.5$):} The primary KAM tori separating the integer resonances are almost  destroyed. The chaotic sea permeates the majority of the phase space, leading to global stochasticity and  chaotic diffusion in momentum space.
    \item   {Fully Developed Chaos ($\delta = 1.0$):} The chaotic sea has overwhelmed almost all surviving island chains. Macroscopic diffusion is  unhindered, permitting particles to rapidly undergo large stochastic energy excursions across the entire phase space. 
    \item   {Intermittent Surfing ($ \delta = 2.0$):} Since $ \delta \gg \delta_{th} \approx 0.25$ (for $\Lambda_0 \sim 1$), the intense wave field  dominates over the background magnetic field. Particles are violently accelerated to ultra-relativistic energies, driving $\Lambda \to 0$. The particle escapes the resonance grid and becomes ponderomotively phase-locked into the macroscopically smooth parabolic pathway, resulting in an large, regular excursion before eventually scattering back into the chaotic sea (L\'evy flight).
\end{enumerate}

\subsection{Lyapunov Spectrum and Onset of Chaos}
To  quantify the onset of stochasticity and validate the transition to chaos, we compute the Lyapunov spectrum across growing laser fields using the Benettin tangent space orthonormalization methodology. The procedure simultaneously integrates the primary three-dimensional nonlinear state trajectory $(\xi, \Lambda, p_z)$ alongside a $3\times3$ matrix of linearized deviation vectors (the tangent space). The tangent vectors are dynamically evolved via the analytic Jacobian matrix $J = \partial \vec{f} / \partial \vec{y}$. To prevent the deviation vectors from numerically collapsing into the single fastest-growing direction, we periodically pause the integration to perform a Gram-Schmidt orthonormalization via QR decomposition. The time-averaged logarithms of the normalization factors (diagonal elements of the upper triangular matrix $R$) directly yield the Lyapunov exponents.

\begin{figure}[ht!]
    \centering
    \includegraphics[width=0.8\textwidth]{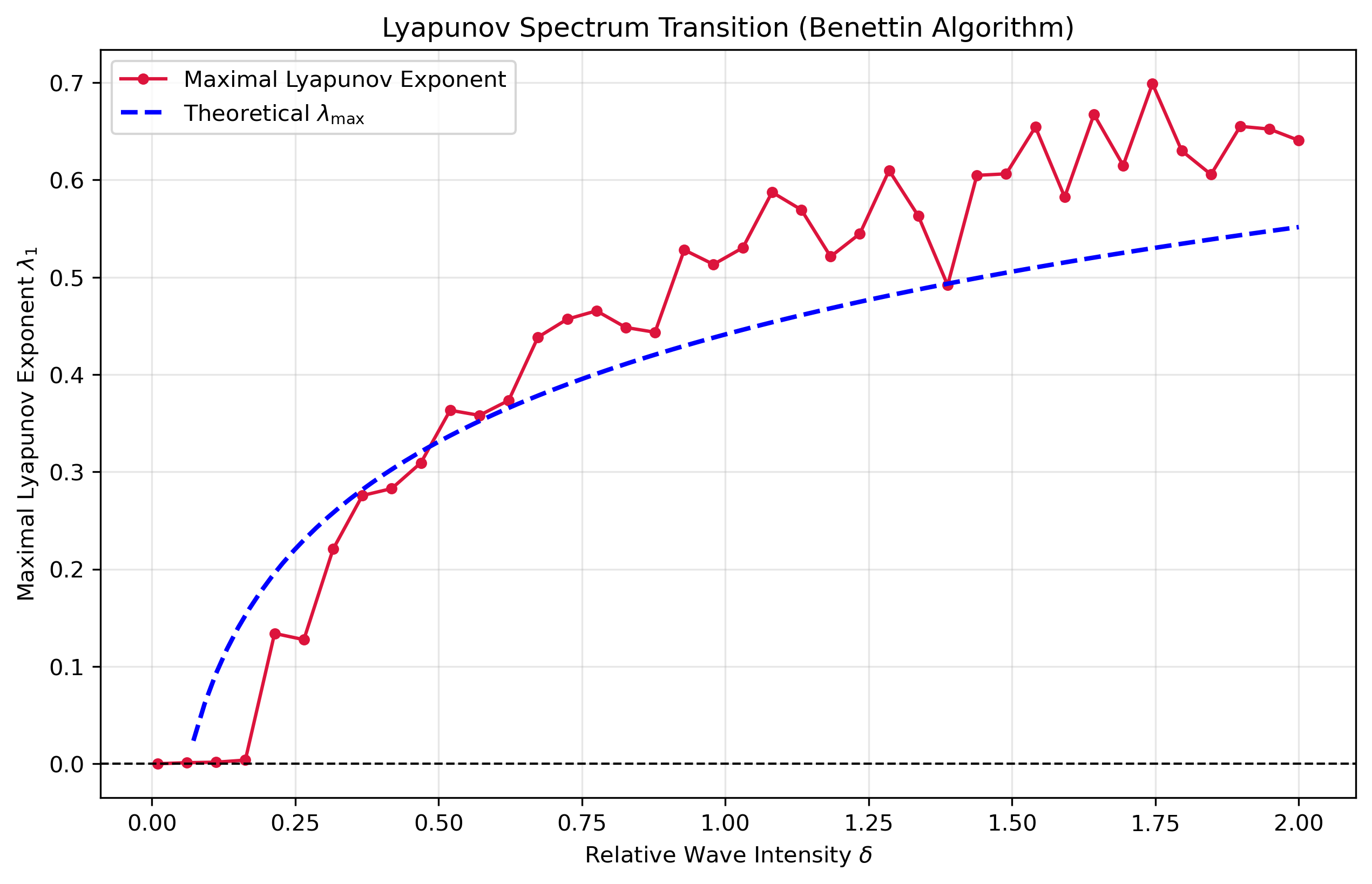}
    \caption{Maximal Lyapunov Exponent (MLE) $\lambda_1$ as a function of the relative wave intensity $ \delta$, calculated via the Benettin algorithm alongside the theoretical maximum $\lambda_{\max}$ curve (Eq. \ref{lambdamax}). The sharp jump from $\lambda_1 = 0$  demarcates the onset of chaos for $ \delta \gtrsim 0.25$.}
    \label{fig:lyapunov}
\end{figure}

The computed Maximal Lyapunov Exponent (MLE)  mirrors the topological transitions observed in the Poincar\'e maps (Figure \ref{fig:lyapunov}). For low wave intensities ($ \delta \lesssim 0.2$), the MLE remains  zero ($\lambda_1 \approx 0.0000$), confirming that the motion is integrable and bounded by stable KAM tori. As $ \delta$ grows past the resonance overlap threshold ($ \delta \gtrsim 0.25$), the MLE sharply jumps to strong positive values ($\lambda_1 > 0.2$), showing  the onset  of global stochasticity. 

As the intensity is increased even further, the Lyapunov exponent does \textit{not} drop back to zero. While intense wave fields drive extreme ponderomotive acceleration and force the particle into macroscopically regular "surfing" pathways ($\Lambda \to 0$), this re-laminarization is never permanent. The background magnetic field $B_0$, no matter how weak relative to the wave, topologically forbids infinite phase-locking; as the particle gains relativistic mass ($\gamma \to \infty$), its effective cyclotron frequency $\omega_c/\gamma \to 0$ drops, but remains non-zero. 
Thus, permanent re-laminarization is topologically forbidden.


If one integrates long enough (much longer than typical numerical cutoffs), the magnetic field slowly but inevitably bends  the particle's trajectory, so that the particle eventually dephases from the wave and  scatters back into the chaotic sea. Because the system exhibits unbounded chaotic diffusion characterized by  intermittent "L\'evy flights" back into the resonance grid, the true asymptotic Lyapunov exponent over infinite time is strictly positive for all $ \delta > \delta_{th}$.

\subsection{Stochastic Diffusion versus Ballistic Surfing}
The coexistence of the chaotic sea and the macroscopically re-laminarized surfing pathways in the high-intensity regime ($\delta \gg 1$) leads to a highly skewed, non-thermal particle energy distribution over finite time intervals. In the globally chaotic regime, almost all particles are accelerated, but the mechanisms and rates of energy gain are  different. The  majority of particles reside within the  chaotic sea, where they undergo stochastic diffusion driven by pseudo-random kicks from overlapping cyclotron resonances. This diffusive heating results in a relatively slow average energy gain characterized by a random walk ($\langle \gamma^2 \rangle \propto t$).

A fraction of particles that are either injected into or chaotically wander into the dynamically sticky boundaries of the accelerator modes become phase-locked with the wave. These few particles undergo the macroscopic intermittent surfing. Because surfing is a ballistic, directed process ($\gamma \propto t$ or $t^2$), these "L\'evy flights" in momentum space generate large  energy gains over the same time interval. Consequently, rather than a uniform acceleration of the particle ensemble, the highly nonlinear phase space dynamics naturally heat the bulk slowly via chaotic diffusion, while  accelerating a select few via ballistic L\'evy flights to form a  high-energy non-thermal tail.

\subsection{Dissipation of wave energy}

This is an important question, as discussed in the Introduction. It should definitely be resolved by heavy duty PIC simulations. Especially demanding are  rare powerful surfing acceleration events. 

Qualitatively, we may identify three energy scales:  (i)  ponderomotive limit 
\be
\gamma_{pond} \sim a_0^2/2
\ee
This is a single-particle limit that follows from the light-cone invariant (see Supplemental Material) for initially stationary particle.

If all the particle experience just the ponderomotive acceleration, the fraction of energy $\eta$ dissipated by the wave is small
\begin{equation}
\eta =   \frac{\omega_p^2}{\omega^2} \ll 1
\end{equation}

(ii)  average \EM\ energy per particle
\be
\gamma_{diff} = a_0^2  \frac{\omega^2}{\om_p^2}
\ee

The ratio
\be
\frac {\gamma_{diff} } {\gamma_{pond} } \approx \frac{\om^2}{\om_p^2} \gg 1
\ee

It is expected that the majority of particles undergoing stochastic diffusion will reach  $\gamma_{diff}$. If {\it all}  the particles achieve  $\gamma_{diff} $, the way will be fully absorbed.

(iii) Surfing limit. A few selected particle will be able to surf the wave. Their energy gain is limited by the total cross-potential associated with the wave, that depends on the lateral extent of the wave, and hence can be related to the total luminosity $L$
\be
\gamma_{surf} \approx \frac{e}{m_e c^{5/2}} \sqrt{L}
\ee

Relative importance of these mechanisms depends on the number of particles participating in each energization channel. Very detailed PIC simulation are required: it's the phases of the particles with respect to waves that are important (again - kinetic,  {\it not } an MHD, questions). Our PIC simulations show that the dissipated fraction remains mild, see   Fig \ref{Total_EM_Energy}.

\section{PIC simulations}
\label{PICs}

The above analysis of the Hamiltonian dynamics of particles in the \EM\ mode is  in the single-particle limit of the collective  plasma phenomena.  We performed  a limited set  of 1D PIC simulations using EPOCH code \cite{2015PPCF...57k3001A}.
Generally, our simulations are consistent with the theory; more detailed investigation of the parameter space will  be done in a  separate publication.

In the simulations, extending over 2 wavelength (periodic boundary condition),  we initialized an X-mode in external \Bf. Initial (dynamic, not external) \EM\ field configuration is
\ba && 
 B_y  =   B_0 (1 + \delta  \sin(k_0 x))
 \nn  && 
  E_z=-  B_0  \delta   \sin(k_0  x)
  \ea
  This is an EM wave of relative amplitude $\delta$ propagating ``to the right''. 
  
   We test the  principled behavior (not trying to cover a wide range in parameter space). The monster shock model predicts dissipation for $\delta \geq 1$. The current model  predicts chaos onset (in the single particle regime) at $\delta   \geq 0.2$.
   
   The parameters of the simulations are as follows.
    Initially particles are cold (not initialized as a consistent nonlinear solution). This should be a reasonable approximation for superluminal waves in high-sigma plasma
  \citep[more detailed nonlinear initialization of subluminal \Alfven waves were done in][]{2026arXiv260501445L}. We verified  that in the proper regime (high-sigma, high frequency $\om \gg \om_p$) fully kinetic simulations closely match single-particle  (option zero$_-$current = T in EPOCH). 
  
   We run a suit of simulations of pair plasma with  $\sigma  \equiv \om_c^2/\om_p^2= 10,100$  (parameter $\sigma$ is defined with respect to each species) and $( \om/\om_p)^2 = 10,100$. All runs have 100 particles per cell, shorter runs have $n_x = 10^5$ cells total, longer ones   $10^4$ cells (hence $5 \times 10^4$ and  $5 \times 10^3$ cells per wavelength correspondingly.

\begin{figure}[ht!]
    \centering
    \includegraphics[width=0.9\textwidth]{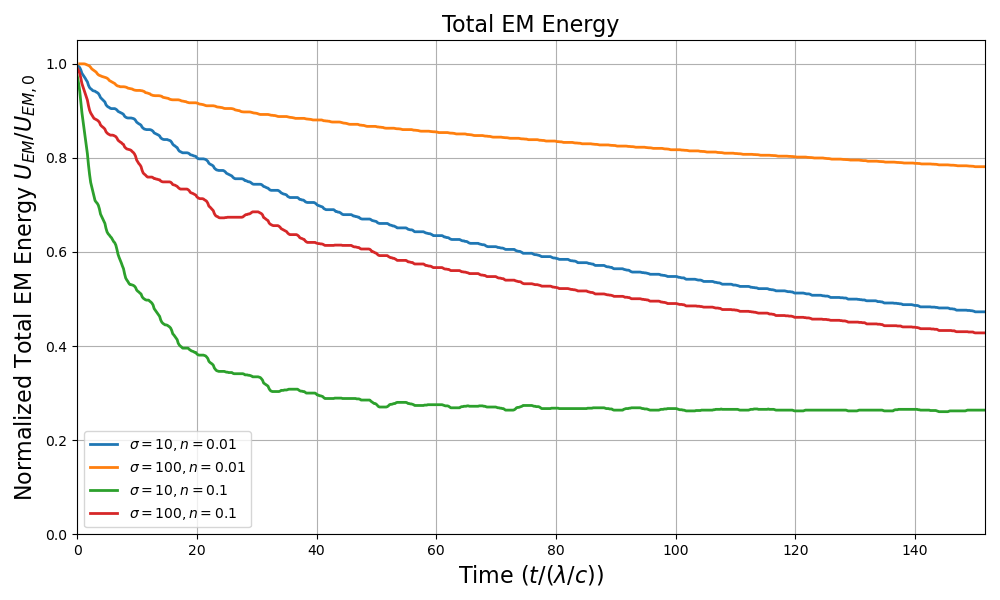}
    \caption{Evolution of the total energy, $\delta= 2$ (notations $n\equiv \om_p^2/\om^2$) .}
    \label{Total_EM_Energy}
\end{figure}

   In Fig. \ref{Total_EM_Energy} we plot evolution of total \EM\ energy for the case $ \delta =2$ (basic configuration with reversing fields of equal absolute amplitude). Though some runs have not achieved asymptotic value, it is clear that a large fraction of the \EM\  survives dissipation.

\begin{figure}[ht!]
    \centering
    \includegraphics[width=0.9\textwidth]{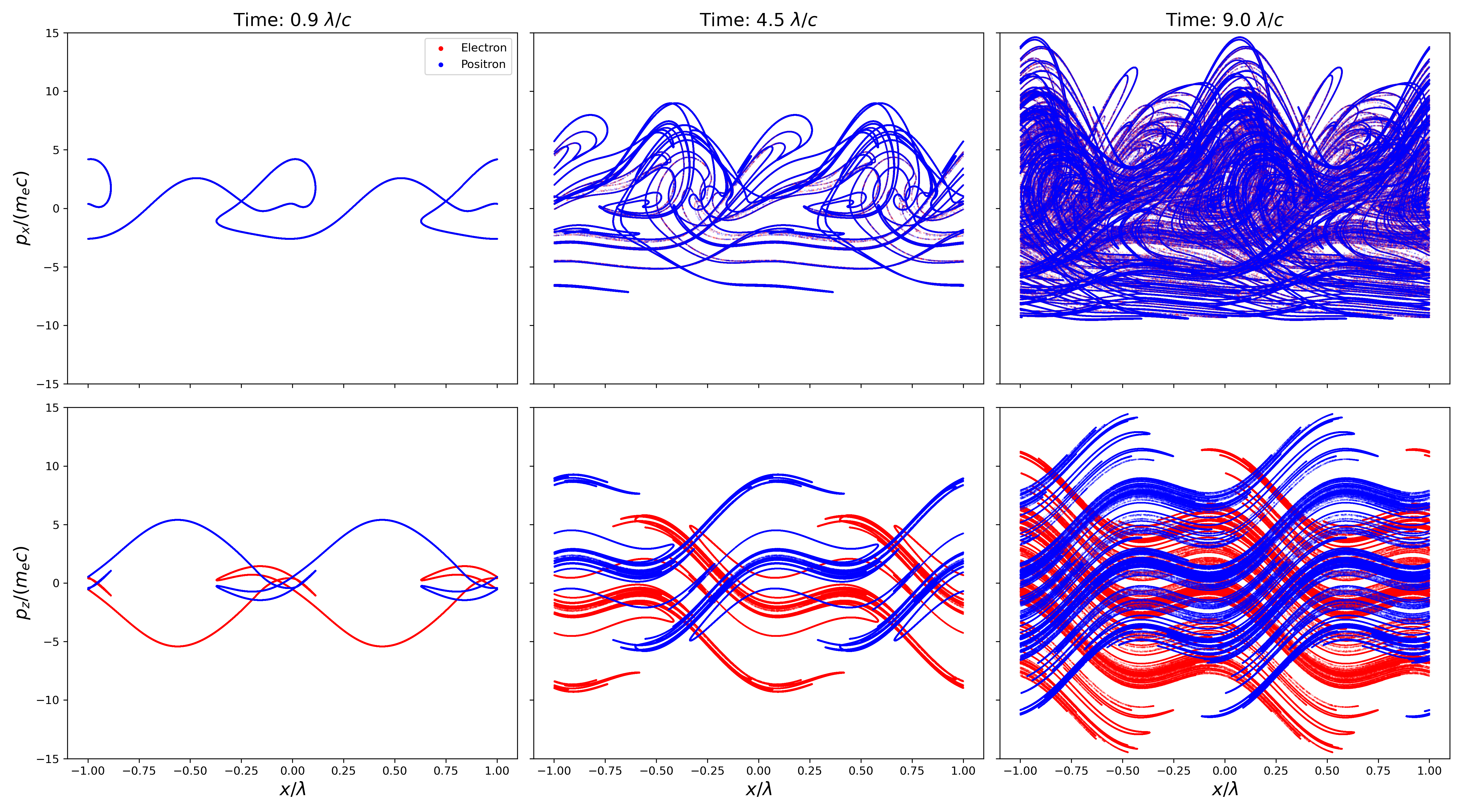}
    \caption{Phase portraits $p_x -x$ and $p_z-x$, $\delta= 2, \sigma =100$. At early times, the $p_x -x$ trajectories for $e^\pm$ match exactly, while  $p_z -x$ trajectories  mirror each other. At later times, due to phase mixing, there is some differences. Also, at  middle times one can  identify trapped particles (island-looking features in top middle plot.}
    \label{Phase_Space_Grid}
\end{figure}

 In Fig. \ref{Phase_Space_Grid} we plot phase portraits $p_x -x$ and $p_z -x$. There are  long-surviving phase correlations, generally consistent with the theoretical  results  (not just MHD bulk acceleration).

\begin{figure}[ht!]
    \centering
    \includegraphics[width=0.9\textwidth]{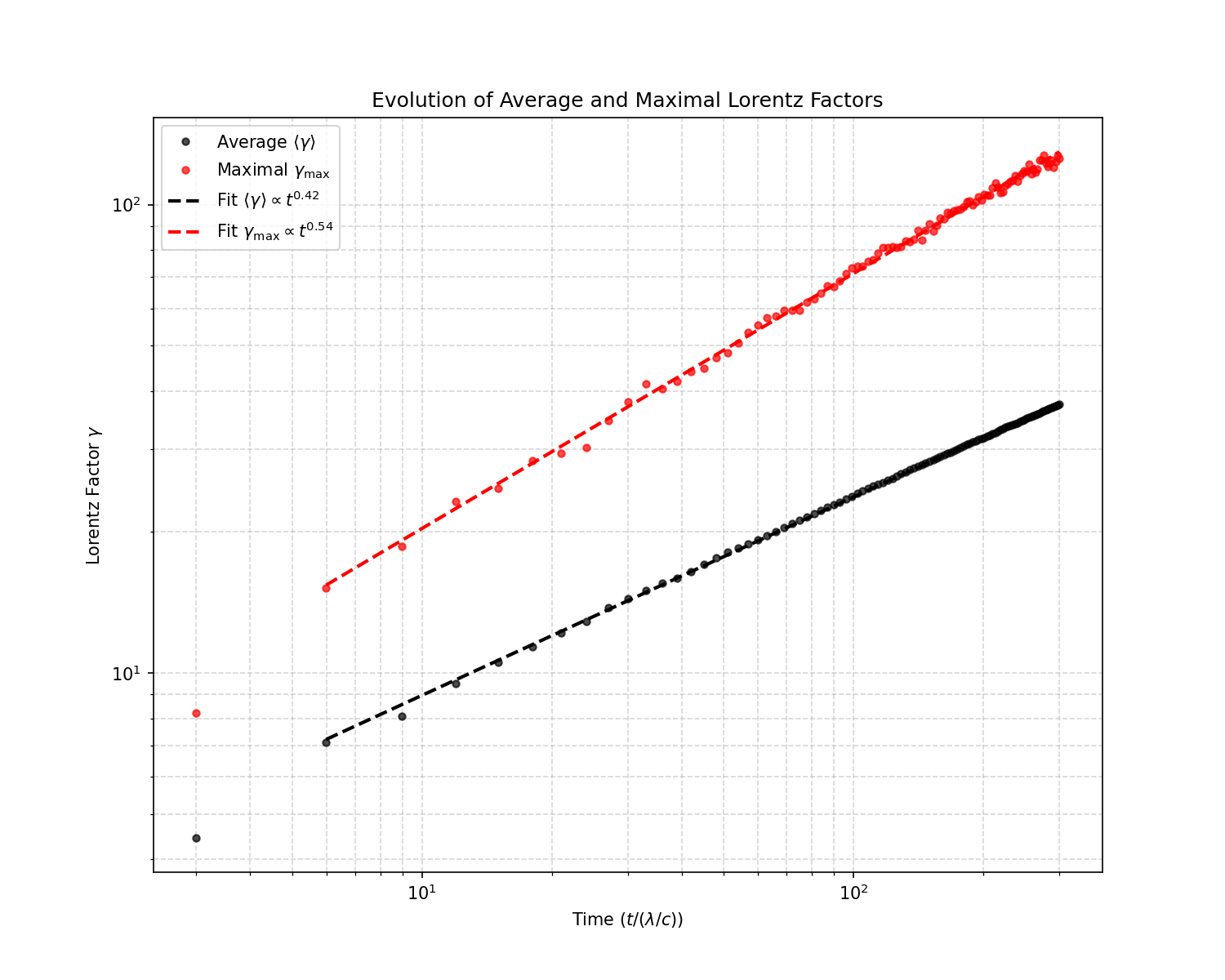}
    \caption{Temporal evolution of the maximal and average {\Lf}s,  $\delta= 2, \sigma =100$  (this is a long simulation to times $300 \lambda/c$ done with smaller resolution $n_x = 500$).}
    \label{gamma_evolution_combined}
\end{figure}

In Fig \ref{gamma_evolution_combined} we plot  temporal evolution of the maximal and average {\Lf}s. The corresponding fits are reasonably close to the expected  random walk scaling of $0.5$; more advanced  simulations are needed to characterize the differences. 

\begin{figure}[ht!]
    \centering
    \includegraphics[width=0.9\textwidth]{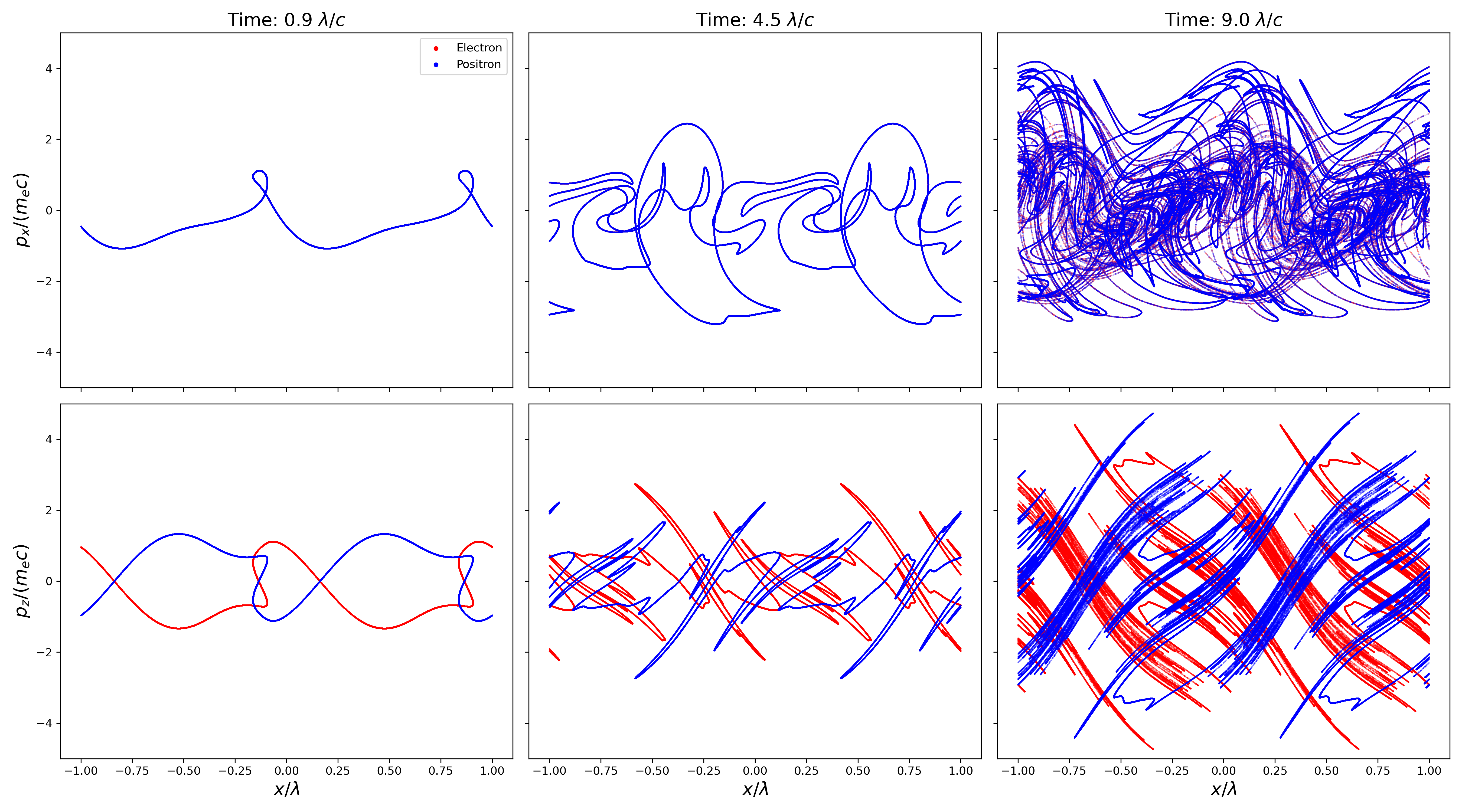}
     \includegraphics[width=0.9\textwidth]{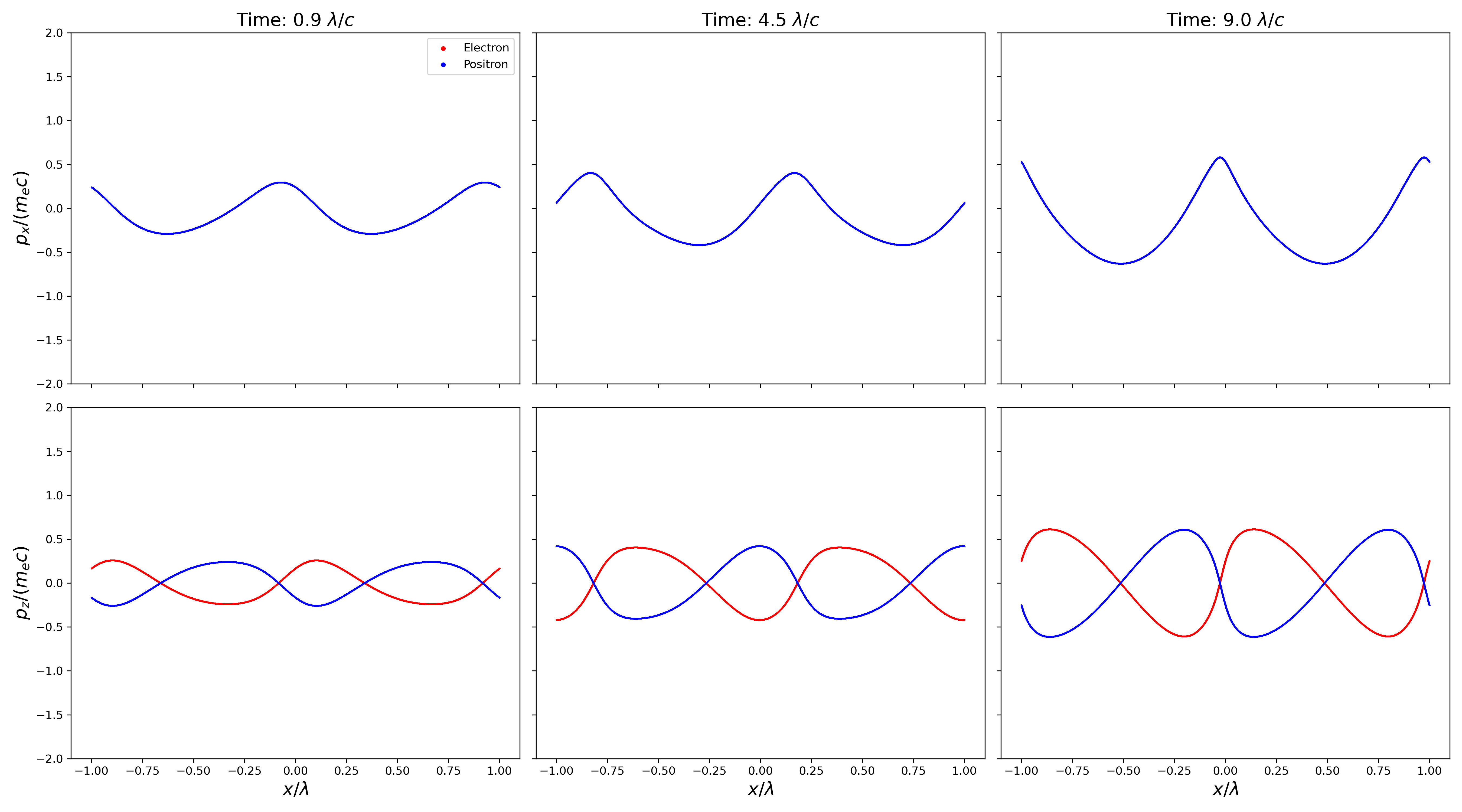}
    \caption{Phase space  for $\delta =0.5$ (top two rows) and  $\delta =0.1$ (bottom two rows). For these parameters  there is no reversal of the \Bf, yet the  $\delta =0.5$ behavior is similar to  $\delta=2$ case, Fig. \ref{Phase_Space_Grid}. The case of $\delta =0.1$ is expected to be subcritical: indeed, we observe regular non-chaotic trajectories.}
    \label{Phase_Space_Grid-delta05}
\end{figure}

Importantly, we also  did runs with  $\delta =0.5$ and $\delta =0.1$  (so, there is no reversal of the \Bf\ - monster shock models would predict no dissipations in this case).  The present work does predict chaos onset for $\delta =0.5$, but not for $\delta =0.1$.  We do indeed observe the expected behavior, Fig. \ref{Phase_Space_Grid-delta05}:   $\delta =0.5$ and $\delta =2$ cases behave similarly chaotic, while   $\delta =0.1$ case is laminar. 

We conclude that PIC simulations seem generally consistent with the theory, but a  more detailed survey of parameters is needed.

\section{Discussion}

In this work, we analyze particle motion in a  nonlinear \EM\ X-modes;  luminal, as well as super- and sub-luminal are considered. The overall dynamics is highly kinetic  (phase-space correlated),  and  cannot be described as  hydrodynamic  flow.

 In the luminal case, we observe chaotization of particles' orbits via  Chirikov resonance overlap mechanism,  and the related  destruction of Kolmogorov-Arnold-Moser (KAM) tori. Chaos onset (non-zero Lyapunov exponent) occurs at $\delta \equiv B_w/B_0 \geq 0.25$  - {\it not} at  $\delta=1$ predicted by the ``monster shock" models. 
 This is confirmed by our PIC simulations.

Intriguingly, in the limit $\delta \gg 1$, when one expects  re-laminarization  of the phase space flow (approaching  the  integrable zero \Bf\ case),  we find that the re-laminarization is never permanent: there is a set of ``surfing particles" that experience large energy  gain via Levi flights, while, the majority experience chaotic ``diffusive heating''. 

There is one  exceptional case: propagation along the guide field. In this particular case trajectories are fully integrable \cite{1964PhRv..135..381R}.  This is due to the fact that for parallel propagation the light-cone variable  $\Lambda$   remains an exact invariant, as noted in Ref. (\cite{2024MNRAS.529.2180L}),  see also Supplemental Material.

Our PIC simulations demonstrate mild  overall wave dissipation. This contradicts   the ``monster shock'' claims of   efficient energy dissipation  \cite{Beloborodov2026}. 
More detailed PIC simulations are needed.

In the highly relativistic regime explored in this work, the mechanisms of trapping (surfing) and intermittent stochastic diffusion become critical. The foundational principles of phase-space trapping within "stochastic webs" and power-law stochastic acceleration in electromagnetic fields were established in Ref. \cite{Zaslavsky1987}. The macroscopic phase-trapping (L\'evy flight) trajectories that we identify in the $\delta \gg 1$ limit are  related to the "surfatron" acceleration mechanism initially proposed  in Ref. \cite{Katsouleas1983},.

The present work, qualitatively, echoes \cite{2025PhRvE.111d5209V}, where chaotic behavior was observed in magnetized shocks. 


Finally, we note an interesting result relegated to the Supplemental Material.
For subluminal waves, at mild intensities,   Cherenkov resonance trapping islands emerge, centered around the phase-matching longitudinal momentum, while at large intensities  $\delta \gg 1 $ Cherenkov islands  overlap with the adjacent regular cyclotron resonance structures, generating chaotic flows similar to the luminal case.

\clearpage
\onecolumngrid
\begin{center}
\textbf{\large Supplemental Material: Particle dynamics in nonlinear electromagnetic waves}
\end{center}
\vspace{2ex}
\setcounter{equation}{0}
\setcounter{figure}{0}
\setcounter{table}{0}
\setcounter{page}{1}
\makeatletter
\renewcommand{\theequation}{S\arabic{equation}}
\renewcommand{\thefigure}{S\arabic{figure}}
\renewcommand{\thetable}{S\arabic{table}}

 \subsection{Perturbation Expansion, Resonance, and Chaotic Limits}

In what follows, we treat the wave field as a perturbation acting on the regular relativistic cyclotron motion induced by guide field.
Unperturbed trajectories exhibit regular gyration with relativistic cyclotron frequency $\Omega_c$. Cyclotron resonance maps to conditions where $\omega \approx n \Omega_c$. The separation between adjacent  cyclotron resonances in energy space is:
\begin{equation}
(\delta \gamma )= \frac{1}{\tilde{\omega}}
\end{equation}
To  derive the resonance width, we expand the locally perturbed system using a pendulum approximation. We transform to the unperturbed action-angle variables $(I, \theta)$, where the action is 
\be
I = \frac{\gamma^2}{2 \om_c}.
\ee
 The unperturbed Hamiltonian is $H_0(I) = \gamma(I)$, yielding the nonlinear cyclotron frequency $\Omega_c $. 

The interaction with the wave creates resonances when the relative phase is stationary: $\dot{\Psi}_n = \omega - n \Omega_c \approx 0$, which defines the resonant energy $\gamma_n = n \om_c / \omega = n / \tilde{\omega}$. Taylor expanding the unperturbed Hamiltonian around the resonant action $I_n$ yields:
\begin{equation}
H_0(I) \approx H_0(I_n) + \Omega_c(I_n) \Delta I + \frac{1}{2} G (\Delta I)^2
\end{equation}
where the nonlinearity parameter (effective inverse mass of the pendulum) is:
\begin{equation}
G = \frac{\partial^2 H_0}{\partial I^2} = \frac{\partial \Omega_c}{\partial \gamma} \frac{\partial \gamma}{\partial I} = \left(-\frac{\om_c }{\gamma_n^2}\right)\left(\frac{\om_c}{\gamma_n}\right) = -\frac{\om_c^2}{\gamma_n^3}
\end{equation}

Isolating the $n$-th harmonic, the locally perturbed system reduces to the standard nonlinear pendulum Hamiltonian:
\begin{equation}
H_{\text{pend}} = \frac{1}{2} G (\Delta I)^2 + V_n \cos(\Psi_n)
\end{equation}
where the effective coupling strength of the $n$-th resonant mode scales as 
\be
V_n \approx  a_0 \tilde{\omega}
\ee

The maximum separatrix width of this pendulum in action space is given by $\Delta I_{\max} = 4 \sqrt{\left| \frac{V_n}{G} \right|}$. Substituting $V_n$ and $G$:
\begin{equation}
\Delta I_{\max} = 4 \sqrt{ a_0 \tilde{\omega} \frac{\gamma_n^3}{\om_c^2} }
\end{equation}
Converting this action width back into energy space using the differential relation $\Delta \gamma = \frac{\partial \gamma}{\partial I} \Delta I = \frac{\om_c}{\gamma_n} \Delta I$, we obtain the localized phase energy width of individual resonance trapping structures:
\begin{equation}
\Delta \gamma_{\text{island}} =  4 \sqrt{a_0 \gamma_n \tilde{\omega}}
\end{equation}

\section{Integrable limits} 
\label{integrable} 

These are   classical \EM\ problem \citep{LLII}, we re-derive them here for consistency.

\subsection{Zero guide field: figure-8}
\label{zeroguide}

In the absence of the background magnetic field ($B_0 = 0$), the right-hand side of the HJ equation becomes independent of $\tau$:
\begin{equation}
2 \omega \frac{\partial \bar{S}}{\partial \xi} \frac{\partial \bar{S}}{\partial \tau} + \left( \frac{\partial \bar{S}}{\partial \tau} \right)^2 = 1    +      a_0^2 \sin^2 \xi.
\end{equation}
We can separate variables by asserting $\bar{S}(\xi, \tau) = S_1(\xi) - \mathcal{E} \tau$, where $\mathcal{E}$ is a separation constant representing a quasi-energy.
\begin{equation}
-2 \omega \mathcal{E} \frac{d S_1}{d \xi} + \mathcal{E}^2 = 1    +      a_0^2 \sin^2 \xi.
\end{equation}
The separation implies:
\begin{equation}
H -  P_x = \mathcal{E} \equiv \text{const}.
\end{equation}
This establishes the conservation of the fundamental light-cone invariant 
\be
\Lambda = \gamma - {p_x}= {\mathcal{E}}
\label{Lambda}
\ee
From the separated HJ equation, we can  solve for the longitudinal canonical momentum $P_x = -k S_1'$ as a function of the phase $\xi$:
\begin{equation}
P_x(\xi) = \frac{1    - \mathcal{E}^2 +      a_0^2 \sin^2 \xi}{2 c \mathcal{E}}.
\end{equation}
Using the equations of motion $v_x =   P_x / H$ and $v_z =   P_z / H$ (where the transverse momentum is $P_z = -e A_z = e   a_0\sin \xi$), we can relate the differential of time $dt$ to the phase $d\xi$ via $dt = \frac{H}{\omega \mathcal{E}} d\xi$. This allows us to analytically integrate the spatial coordinates with respect to $\xi$:
\begin{align}
z(\xi) &= \int \frac{  P_z}{\omega \mathcal{E}} d\xi = - \frac{ a_0 }{\omega \mathcal{E}} \cos \xi, \\
x(\xi) &= \int \frac{  P_x}{\omega \mathcal{E}} d\xi = \tilde{v}_d \frac{\xi}{\omega} - \frac{    a_0^2 }{8 \omega \mathcal{E}^2} \sin 2\xi,
\\
\tilde{v}_d  & = \frac{1}{2 \mathcal{E}^2} \left( 1    - \mathcal{E}^2 + \frac{    a_0^2 }{2 } \right)
\end{align}
($\tilde{v}_d$ characterizes the steady longitudinal ponderomotive drift). 
These  parametric equations describe a "Figure-8" trajectory in the average rest frame (where the drift vanishes), consisting of transverse quivering at frequency $\omega$ and longitudinal oscillation at frequency $2\omega$. 

\subsection{Cyclotron motion in a constant magnetic field via Hamilton-Jacobi approach}
\label{app:cyclotron_HJ}

In a constant, uniform magnetic field $\mathbf{B} = B_0 \hat{\mathbf{y}}$, using the Landau gauge $\mathbf{A} = -B_0 x \hat{\mathbf{z}}$, the Hamiltonian $H$ becomes independent of time $t$ and the transverse coordinates $y$ and $z$:
\begin{equation}
H = \sqrt{1 + P_x^2 + P_y^2 + (P_z + B_0 x)^2}.
\end{equation}
We can separate variables by asserting $\bar{S}(t, x, y, z) = S_1(x) + P_y y + P_z z - \mathcal{E} t$, where $\mathcal{E}$ is a separation constant representing the conserved energy, and $P_y, P_z$ are the conserved transverse canonical momenta. 

From the separated HJ equation, $H = \mathcal{E}$, we can  solve for the longitudinal canonical momentum $P_x = S_1'$ as a function of $x$:
\begin{equation}
P_x(x) = \pm \sqrt{\mathcal{E}_\perp^2 - (P_z + B_0 x)^2},
\label{eq:Sx_deriv}
\end{equation}
where we have defined the constant transverse energy squared as $\mathcal{E}_\perp^2 \equiv \mathcal{E}^2 - 1 - P_y^2$. Integrating Eq. (\ref{eq:Sx_deriv}) yields $S_1(x)$:
\begin{equation}
S_1(x) = \pm \int \sqrt{\mathcal{E}_\perp^2 - (P_z + B_0 x)^2} \, dx.
\end{equation}

Using Jacobi's theorem, we differentiate the total action $\bar{S}$ with respect to the constant momenta $\mathcal{E}$, $P_y$, and $P_z$. This allows us to analytically integrate the spatial coordinates:
\begin{align}
\frac{\partial \bar{S}}{\partial \mathcal{E}} &= -t + \frac{\partial S_1}{\partial \mathcal{E}} = \alpha_t, \label{eq:jacobi_t} \\
\frac{\partial \bar{S}}{\partial P_y} &= y + \frac{\partial S_1}{\partial P_y} = \alpha_y, \label{eq:jacobi_y} \\
\frac{\partial \bar{S}}{\partial P_z} &= z + \frac{\partial S_1}{\partial P_z} = \alpha_z. \label{eq:jacobi_z}
\end{align}
Here, we  use the separated form of the total action:
\begin{equation}
\bar{S}(t, x, y, z) = S_1(x) + P_y y + P_z z - \mathcal{E} t.
\end{equation}

Evaluating the derivative in Eq. (\ref{eq:jacobi_z}) yields the trajectory in the $x-z$ plane:
\begin{equation}
z - \alpha_z = \mp \int \frac{P_z + B_0 x}{\sqrt{\mathcal{E}_\perp^2 - (P_z + B_0 x)^2}} \, dx.
\end{equation}
Letting $u = P_z + B_0 x$, the integral evaluates to $\pm \frac{1}{B_0} \sqrt{\mathcal{E}_\perp^2 - u^2}$. Squaring both sides and substituting $u$ back yields the geometric equation of the orbit:
\begin{equation}
(z - \alpha_z)^2 + \left( x + \frac{P_z}{B_0} \right)^2 = \frac{\mathcal{E}_\perp^2}{B_0^2}.
\end{equation}
This parametric equation describes a circular trajectory in the $x-z$ plane centered at $(x_c, z_c) = (-P_z / B_0, \alpha_z)$ with Larmor radius $r_L = \mathcal{E}_\perp / B_0$. 

To find the time evolution, we evaluate Eq. (\ref{eq:jacobi_t}):
\begin{equation}
t + \alpha_t = \pm \mathcal{E} \int \frac{dx}{\sqrt{\mathcal{E}_\perp^2 - (P_z + B_0 x)^2}}.
\end{equation}
Using the substitution $P_z + B_0 x = -\mathcal{E}_\perp \cos(\Phi)$, this gives $t + \alpha_t = \frac{\mathcal{E}}{B_0} \Phi$. Thus, the phase angle evolves linearly with time as $\Phi(t) = \Omega_c (t + \alpha_t)$, where $\Omega_c = B_0 / \mathcal{E}$ is the relativistic cyclotron frequency.

\section{Superluminal Regime ($n< 1$)}
\label{n1}

In the absence of guide field, this case  can  be studies by applying  a Lorentz transformation to Clemmow \cite{1974JPlPh..12..297C}  reference frame moving at velocity $V =   1/ v_{ph} < c$ along the $x$-axis. In the  Clemmow frame,  the wave field is  spatially uniform and purely time-dependent  oscillating homogeneous electric field. The particle undergoes secular transverse oscillations without the secular longitudinal ponderomotive acceleration seen in the $\beta=1$ case.

With an external magnetic field $B_0$, transformed to the $k'=0$ frame, the particle experiences a constant magnetic field and an oscillating uniform electric field. If the effective driving frequency matches the relativistic cyclotron frequency, classical cyclotron resonance occurs. Crucially, because $\beta > 1$, the Cherenkov resonance condition $v_x = v_{ph}$ is physically unattainable ($v_x \le c < v_{ph}$), precluding wave-trapping phenomena.

To  confirm the structural stability of the superluminal regime, we numerically evaluate the stroboscopic proper-time phase intersections for a refractive index of $n=0.5$ (where the phase velocity is strictly superluminal, $v_{ph} = 2c$). 

\begin{figure}[ht!]
    \centering
    \includegraphics[width=0.9\textwidth]{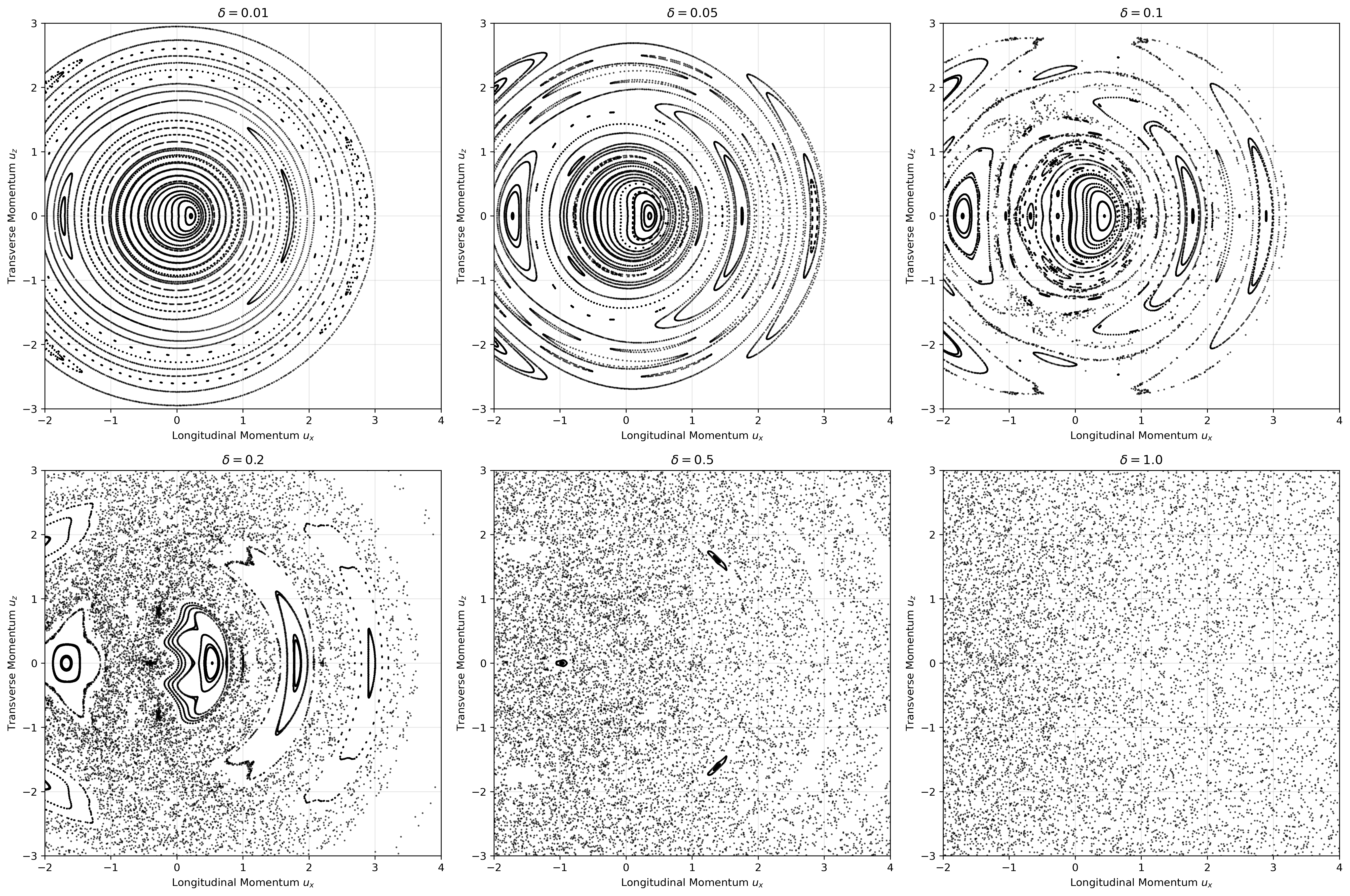}
    \caption{Poincar\'e maps illustrating the superluminal phase space topology for $n=0.5$ ($v_{ph} = 2c$). The complete absence of physically accessible Cherenkov resonance trapping allows stable invariant KAM curves to dominate the phase space even at large wave intensities, severely suppressing the onset of global chaos.}
    \label{fig:poincare_n05}
\end{figure}

Figure \ref{fig:poincare_n05} presents the resulting topological maps across varying wave intensities. Unlike the subluminal case, the phase space  lacks  Cherenkov trapping islands because the resonance condition ($v_x = v_{ph}$) is strictly prohibited by special relativity ($v_x \le c < 2c$). Consequently, even at relatively large wave intensities ($\delta \sim 1.0$), the phase space remains predominantly regular, characterized by robust, unbroken invariant KAM tori. Global stochasticity is exceptionally suppressed, confirming that superluminal non-autonomous electromagnetic waves cannot trap particles and thereby fail to induce the  chaotic overlap required for surfing acceleration.

\section{Subluminal Regime ($n >  1$)}
\label{n2}
When the phase velocity is subluminal ($\omega <   k$), the dynamics fundamentally shift because particles can surf the wave/ experience Cherenkov resonance. 

In the absence of guide field, we can perform a Lorentz transformation to the rest frame of the wave, moving at $V = v_{ph} = \beta c < c$. In this frame, $\omega' = 0$, and the wave manifests as a static, spatially periodic magnetic structure (a magnetic wiggler). Particles can become trapped in the stationary potential wells of the wave if their energy is sufficiently low, or they execute unbound, modulated streaming motion if their energy is high. 

The presence of $B_0$ in the subluminal regime enables true wave-particle trapping via the Cherenkov resonance, $v_x = v_{ph}$. A particle near this velocity remains in constant phase with the wave, allowing continuous energy exchange. The overlap of the Cherenkov resonance island with adjacent cyclotron resonance islands severely lowers the stochasticity threshold. The phase space exhibits complex topology, featuring deep trapping regions interspersed with a highly chaotic sea resulting from the overlapping resonances, leading to phenomena like surfing acceleration.

When the refractive index $n > 1$ (subluminal phase velocity, $v_{ph}/c = 1/n < 1$), the light-cone momentum $\Lambda = \gamma - p_x$ derived for the luminous case ($n=1$) is no longer a strict invariant. Consequently, integrating the non-autonomous dynamical system natively over the wave phase $\xi$ becomes structurally ill-posed for wave-trapped particles, as $d\xi/ds$ can cross zero and flip signs. 
Instead, we numerically integrate the fully generalized proper-time ordinary differential equations and dynamically locate the stroboscopic phase intersections ($\xi \mod 2\pi = 0$) using root-finding event detection over  trajectory ensembles.

\begin{figure}[ht!]
    \centering
    \includegraphics[width=0.9\textwidth]{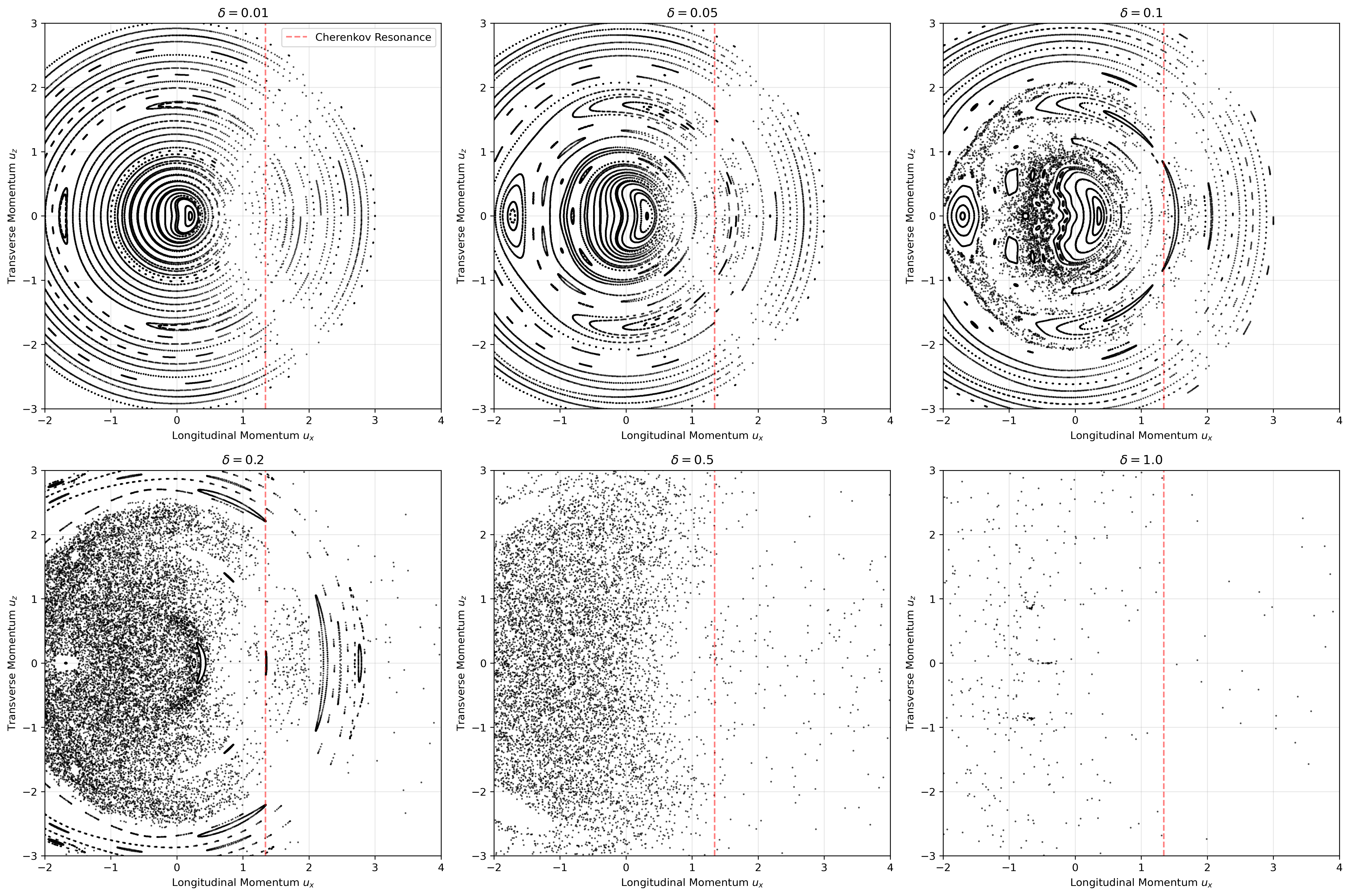}
    \caption{Poincar\'e maps illustrating the subluminal phase space topology for $n=1.5$. The vertical red dashed line indicates the theoretical Cherenkov resonance condition. As the wave intensity $\delta$ increases, the trapping islands expand, severely overlap, and rapidly deteriorate into a global stochastic sea.}
    \label{fig:poincare_n15}
\end{figure}

Figure \ref{fig:poincare_n15} illustrates the topological evolution of the subluminal phase space for $n=1.5$. At lower wave intensities, clear Cherenkov resonance trapping islands emerge, centered around the phase-matching longitudinal momentum where particles surf the wave at $v_x = c/n$. As the relative wave intensity $\delta$ increases, these  Cherenkov islands violently overlap with the adjacent regular cyclotron resonance structures. This rapidly lowers the macroscopic stochasticity threshold and plunges the phase space into  interconnected chaotic sea long before comparable intensities in the $n=1$ case.

Finally, we note that an  anomalous cyclotron resonance is not achievable in the particular geometry. 
The external magnetic field $\mathbf{B}_0 = B_0 \hat{y}$ is strictly perpendicular to the wave propagation vector $\mathbf{k} = k \hat{x}$. Consequently, the unperturbed cyclotron motion occurs entirely in the $x$-$z$ plane, with the longitudinal velocity oscillating as $v_x(t) \propto \cos(\Omega_c t)$. Because there is no guiding-center drift along the $\hat{x}$ direction, the time-averaged longitudinal velocity is identically zero ($\langle v_x \rangle = 0$). 

While a particle may \textit{momentarily} outrun the wave during a fraction of its cyclotron orbit, it is structurally impossible for it to sustain the secular negative Doppler shift required to drive an anomalous resonance. Therefore, despite the wave being subluminal, the purely transverse magnetic geometry structurally  suppresses anomalous cyclotron resonances, restricting the wave-particle interactions exclusively to the Cherenkov resonance ($\langle v_x \rangle \approx v_{ph}$ during trapping) and the normal cyclotron harmonics ($\omega = n \Omega_c$).

\section{Complementary Expansion: Wave-Dominated Regime ($ \delta \gg 1 $)}
\label{Complementary}

To study the complementary limit where the external guide field acts as a small perturbation on the nonlinear wave dynamics (expansion near $ 1/\delta \to 0 $), it is convenient to re-scale the proper time by the wave field's characteristic frequency $\omega_w = B_w = \delta \omega_c$. We define a new dimensionless proper time $\chi = \delta s$ (By simply stretching the timeline by a factor of  $\delta$, the system  shifts from the perspective of the background magnetic field (using 
$s$) to the perspective of the intense electromagnetic wave (using  $\chi$).  The derivatives transform as $\frac{d}{ds} = \delta \frac{d}{d\chi}$. 

Substituting this into the dynamical system (\ref{Main1}), we obtain:
\begin{align}
\frac{d\xi}{d\chi} &= \frac{\tilde{\omega}}{\delta} \Lambda = \frac{1}{a_0} \Lambda, \\
\frac{d\Lambda}{d\chi} &= \frac{1}{\delta} p_z, \\
\frac{dp_z}{d\chi} &= \Lambda \cos\xi + \frac{1}{\delta} \frac{1 - \Lambda^2 + p_z^2}{2\Lambda}. 
\end{align}

In the  unperturbed pure-wave limit ($ 1/\delta = 0 $), the light-cone variable $\Lambda$ becomes a constant of motion ($\Lambda = \Lambda_0$). Furthermore, $dp_z / d\chi = \Lambda_0 \cos\xi = a_0 \cos\xi (d\xi/d\chi)$, revealing a second  invariant corresponding to the canonical transverse momentum: $P_z = p_z - a_0 \sin\xi = \text{const}$. 

Transforming the full system to use these unperturbed invariants ($\Lambda, P_z$)  isolates the slow guide field perturbation:
\begin{align}
\frac{d\xi}{d\chi} &= \frac{1}{a_0} \Lambda, \\
\frac{d\Lambda}{d\chi} &= \boxed{ \frac{1}{\delta} (P_z + a_0 \sin\xi) }, \\
\frac{dP_z}{d\chi} &= \boxed{ \frac{1}{\delta} \times  \frac{1 - \Lambda^2 + (P_z + a_0 \sin\xi)^2}{2\Lambda} }.
\label{Main2}
\end{align}
This system is the  analogue to (\ref{Main1}). It demonstrates that for $1/\delta = 0$, the motion is perfectly integrable (free-streaming in phase $\xi$ with constant invariants), while the finite $1/\delta$ boxed terms introduce slow, symmetry-breaking drifts that destroy the invariant tori from the strong-wave side.


The system (\ref{Main2}) naturally reveals the mechanisms of trapping and chaos in the high-intensity regime. Taking the second derivative of the wave phase yields a pendulum equation driven by a slowly varying torque:
\begin{equation}
\frac{d^2\xi}{d\chi^2} = \frac{1}{a_0} \frac{d\Lambda}{d\chi} = \frac{1}{\delta} \sin\xi + \frac{P_z}{a_0 \delta}.
\end{equation}
Near the main surfing resonance ($P_z \approx 0$), this describes a standard non-linear pendulum $d^2\xi/d\chi^2 - \delta^{-1} \sin\xi = 0$. The full separatrix width (trapping island) of this pendulum in the phase velocity $d\xi/d\chi$ is $4/\sqrt{\delta}$. Using $\Lambda = a_0 (d\xi/d\chi)$, the momentum width of the primary trapping resonance is:
\begin{equation}
\Delta \Lambda_{\text{island}} = 4 a_0 \sqrt{\frac{1}{\delta}}.
\end{equation}

The weak guide field causes the transverse momentum $P_z$ to undergo slow macroscopic Larmor modulation. Linearizing the slow system reveals this modulation frequency in $\chi$-time to be $\Omega_{\text{slow}} = 1/\delta$. This low-frequency perturbation creates an infinite ladder of sideband resonances (which  correspond  to the cyclotron resonances $\Lambda_m = m/\tilde{\omega}$) separated in momentum by:
\begin{equation}
(\delta \Lambda) = a_0 \Omega_{\text{slow}} = \frac{a_0}{\delta}.
\end{equation}

The Chirikov resonance overlap parameter $K$ for this strong-wave expansion is the ratio of the primary island width to the sideband separation:
\begin{equation}
K = \frac{\Delta \Lambda_{\text{island}}}{(\delta \Lambda)} = \frac{4 a_0 / \sqrt{\delta}}{a_0 / \delta} = 4 \sqrt{\delta}.
\end{equation}
The onset of global stochasticity ($K \ge 1$) therefore occurs precisely at:
\begin{equation}
4 \sqrt{\delta} \ge 1 \implies \delta \ge \frac{1}{16}.
\end{equation}
This  recovers the identical chaos threshold derived from the weak-wave limit.

Finally, deep in the chaotic regime ($K \gg 1$), the discrete standard map trajectory divergence per slow modulation step $T_{\chi} = 2\pi / \Omega_{\text{slow}} = 2\pi \delta$ is $\lambda_{\text{step}} \approx \ln(K/2) = \ln(2\sqrt{\delta})$. The continuous Lyapunov exponent with respect to the physical wave phase $\xi$ (spanning $\Delta\xi = 2\pi \tilde{\omega}$ per step) is therefore:
\begin{equation}
\lambda_{\max} \approx \frac{\lambda_{\text{step}}}{\Delta\xi} = \frac{1}{2\pi \tilde{\omega}} \ln\left(2\sqrt{\delta}\right).
\end{equation}
This  confirms that in the wave-dominated limit, the maximal Lyapunov exponent continues to increase logarithmically with the relative wave intensity $\delta$,  as observed numerically in Fig. \ref{fig:lyapunov}.

\section{Integrability of the Purely Longitudinal Guide Field}
\label{Longitudinal}


If the external guide field is purely longitudinal, $\mathbf{B}_{ext} = B_0 \hat{\mathbf{x}}$, the Lorentz force equations evaluated in proper time $s$ become:
\begin{align}
\frac{dp_x}{ds} &= -p_z B_w \cos\xi \label{eq:long_px} \\
\frac{dp_y}{ds} &= p_z B_0 \label{eq:long_py} \\
\frac{dp_z}{ds} &= \gamma E_w \cos\xi - p_x B_w \cos\xi - p_y B_0
\end{align}
The energy equation is:
\begin{equation}
\frac{d\gamma}{ds} = p_z E_w \cos\xi \label{eq:long_gamma}
\end{equation}

Subtracting Eq. (\ref{eq:long_px}) from Eq. (\ref{eq:long_gamma}) and recalling $E_w = B_w$ reveals that the light-cone variable $\Lambda = \gamma - p_x$ is strictly conserved:
\begin{equation}
\frac{d\Lambda}{ds} = \frac{d\gamma}{ds} - \frac{dp_x}{ds} = 0 \implies \Lambda = \text{const}.
\end{equation}
Because $\Lambda$ is a constant of motion, the wave phase $\xi$ evolves strictly linearly in proper time:
\begin{equation}
\frac{d\xi}{ds} = \omega \gamma - k p_x = \omega (\gamma - p_x) = \omega \Lambda = \text{const} \implies \xi(s) = \omega \Lambda s + \xi_0
\end{equation}
Substituting $\Lambda$ into the $p_z$ equation yields:
\begin{equation}
\frac{dp_z}{ds} = (\gamma - p_x) E_w \cos\xi - p_y B_0 = \Lambda E_w \cos(\omega \Lambda s + \xi_0) - p_y B_0
\end{equation}

The transverse dynamics are therefore governed by the closed 2D system:
\begin{align}
\frac{dp_y}{ds} &= B_0 p_z \\
\frac{dp_z}{ds} &= -B_0 p_y + \Lambda E_w \cos(\omega \Lambda s + \xi_0)
\end{align}
Differentiating $p_z$ with respect to $s$ produces the equation of a driven linear harmonic oscillator:
\begin{equation}
\frac{d^2 p_z}{ds^2} + B_0^2 p_z = -\Lambda^2 \omega E_w \sin(\omega \Lambda s + \xi_0)
\end{equation}
Because the system reduces exactly to a linear harmonic oscillator in the transverse plane, it is {exactly integrable} for any arbitrary wave amplitude $E_w$. The particle trajectory is simply a superposition of the natural cyclotron motion (at frequency $B_0\equiv \om_c$) and the harmonic driven response (at frequency $\omega \Lambda$).


\bibliographystyle{apsrev4-2}
\bibliography{references}
\end{document}